# Stochastic spin-orbit-torque synapse and its application in uncertainty quantification

Cen Wang[1#], Guang Zeng[2#], Xinyu Wen[1], Yuhui He[1], Wei Luo[1], Shiwei Chen[2], Shiheng Liang[2*], Yue Zhang[1*]

1. School of Integrated Circuits, Huazhong University of Science and Technology, Wuhan, 430074, P. R. China

2. School of Physics, Hubei University, Wuhan, 430062, P. R. China

#Cen Wang and Guang Zeng equally contribute to this work.

*Corresponding authors: yue-zhang@hust.edu.cn (Yue Zhang); shihengliang@hubu.edu.cn (Shiheng Liang)

**Abstract**

Stochasticity plays a significant role in the low-power operation of a biological neural network. In an artificial neural network (ANN), stochasticity also contributes to critical functions such as the uncertainty quantification (UQ) for estimating the probability for the correctness of prediction. This UQ is vital for cutting-edge applications, including medical diagnostics, autopilots, and large language models. Thanks to high computing velocity and low dissipation, a spin-orbit-torque (SOT) device exhibits significant potential for implementing the UQ. However, up until now, the application of UQ for stochastic SOT devices remains unexplored. In this study, based on SOT-induced stochastic magnetic domain wall (DW) motion with varying velocity, we fabricated an SOT synapse that could emulate stochastic weight update following the Spike-Timing-Dependent-Plasticity (STDP) rule. Furthermore, we set up a stochastic Spiking-Neural-Network (SNN), which, when compared to its deterministic counterpart, demonstrates a clear advantage in quantifying uncertainty for diagnosing the type of breast tumor (benign or malignant).

**Introduction**

Stochasticity plays a significantly crucial role in the low-power operation of a biological neural network (BNN). Beyond the stochastic firing of neurons, synapses responsible for modulating the strength of connections between pre-neurons and post-neurons also manifest non-deterministic behaviors. The random events associated with a biological synapse stem from the spontaneous opening of intracellular $Ca^{2+}$ stores, synaptic $Ca^{2+}$ channel noise, and the random positioning of vesicles with a broad size distribution [1]. This stochasticity is discernable through distinct outputs under identical input stimulation [Fig. 1(a)].

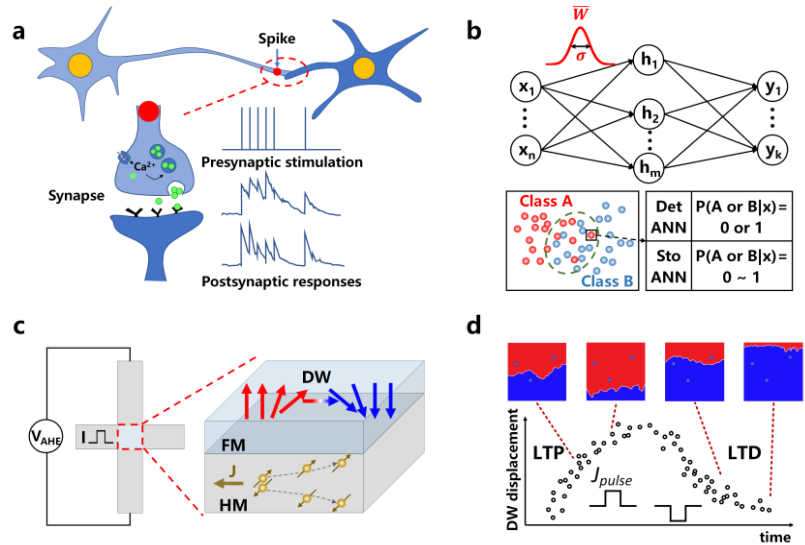

**Figure 1. (a) Illustration depicting a biological synapse facilitating communication between pre-synaptic and post-synaptic neurons. The stochasticity manifests as different postsynaptic outputs when subjected to identical presynaptic input stimulations. This non-deterministic synaptic behavior arises from the spontaneous opening of intracellular $Ca^{2+}$ stores, synaptic $Ca^{2+}$ channel noise, and the random distribution of vesicles with a wide size range. (b) The stochastic weight update in an ANN can be harnessed to implement UQ, a crucial aspect for estimating uncertainty in classifying diverse objects with mixed features. (c) Within an SOT device, injecting a current pulse induces the variation of $R_{AHE}$, achieved through the DW motion driven by SOT. (d) The DW velocity exhibits continuous variation during its motion, encompassing rapid DW motion immediately after depinning and slower motion as the DW approaches the border. This behavior can emulate the processes of Long-Term Potential (LTP) and Long-Term Depression (LTD) in the nonlinear weight update.**

As inspired by BNN, the Artificial Neural Network (ANN) has been widely harnessed to execute various functions in artificial intelligence, including recognition, analysis, and inference. In a deterministic ANN, the presence of noise in an electronic device could be detrimental for the accuracy of predictions. However, judiciously leveraging noise can offer crucial functionalities that a conventional ANN lacks. For example, moderate noise can expedite the training process of an

ANN and be utilized for probabilistic computing [2]. In addition to that, grounded in Bayesian inference, controlled noise also plays a pivotal role in UQ [3,4], which includes the calculation of posterior probability based on observational outcomes and the estimation of uncertainty in predictions [Fig. 1(b)]. This UQ hold particular significance for cutting-edge applications such as medical diagnostics, autopilots, and large language models [5-7].

Among various types of ANNs, the Spiking Neural Network (SNN) holds unique potential for implementing UQ because the presence or absence of spikes in an SNN correlated with the sampling from a binary random variable [8]. On the other hand, the event-driven and asynchronous spiked-based parallel processing results in low power consumption [9], a crucial aspect for UQ demanding substantial computational effort. To date, there has been widespread demonstration of the application of UQ using SNNs with stochastic neurons [10, 11]. In addition to neurons, stochastic synapses also play a critical role in UQ. For example, it has been theoretically established that spiking artificial neurons in a noisy synapse environment can perform Bayesian inference based on incomplete observations [12]. Stochastic synapses have also found application in neural sampling machines for approximating Bayesian inference through Monte Carlo sampling [13].

Different electronic devices, such as memresistors, ferroelectrics, and spintronics devices, can be exploited to emulate an artificial synapse in an ANN. In the Spin-Orbit-Torque (SOT) device, current pulses can induce a continuous variation of anomalous Hall resistance ($R_{AHE}$) characterized by ultra-low dissipation (fJ/bit) and rapid processing (ps) [Fig. 1(c)] [14,15]. Consequently, the SOT device holds promise for stochastic neuromorphic computing which requires computational power with low energy consumption. Despite increased attention to the stochasticity of SOT devices in recent years [16,17], the exploration of their application in UQ remains largely uncharted.

From a microscopic perspective, the modulation of $R_{AHE}$ in an SOT device is based on current-driven domain wall (DW) motion [Fig. 1(c)], including a linear change in $R_{AHE}$ when a DW moves at a uniform velocity and a nonlinear variation of $R_{AHE}$ when DW motion features variable speeds [18,19]. Inducing a uniform DW motion necessitates driving the DW into a narrow strip. However, in more general scenario, DW motion comprises distinct stages with varying velocities, including the initial post-depinning fast motion and the subsequent slower motion as the DW approaches the edge [Fig. 1(d)]. This DW motion with diverse velocities could lead to nonlinearly varied $R_{AHE}$, which corresponds to nonlinear weight update for a synapse in an SNN. Furthermore, owing to thermal fluctuation and random distribution of pinning centers, the DW motion also exhibits stochastic behaviors [20,21], contributing to stochasticity in weight updates.

In this paper, leveraging SOT-induced DW motion with varying velocities, we designed and fabricated a stochastic synapse to emulate nonlinear weight updates in an SNN. Employing this SOT synapse, we established a stochastic SNN for the classification of breast tumors, assessing the uncertainty of the prediction. In contrast to a deterministic neural network, the SNN equipped with the stochastic SOT synapse adeptly gauges the uncertainty for the prediction outcomes.

**Results and discussions**

To determine the optimal structure of an SOT device suitable for nonlinear weight update, we first simulated the SOT-induced variation of the z-component of magnetization ($m_z$) within the cross regions of four Hall bars with different shapes by using the micromagentic simulation [Fig.2a] (Refer to the "Methods" section for more details). This $m_z$ variation arises from the DW motion shown in Fig.3. In the Hall bars I and II, the length ($L$) of the cross region is much smaller than the width ($W$), with consideration given to the edge roughness in structure II. A square-shaped cross-region, where $L$ equals $W$, was assumed in the Hall bar III, while $L > W$ in Hall bar IV. The simulation results demonstrated a linear $m_z$ variation attributed to uniform DW motion in Hall bar IV, whereas nonlinear $m_z$ variation was obvious for the Hall bars I and II. Especially, the $m_z$ variation in the Hall bar II closely resembles the procedures of Long-Term Potential (LTP) and Long-Term Depression (LTD) associated with the nonlinear weight update in an SNN.

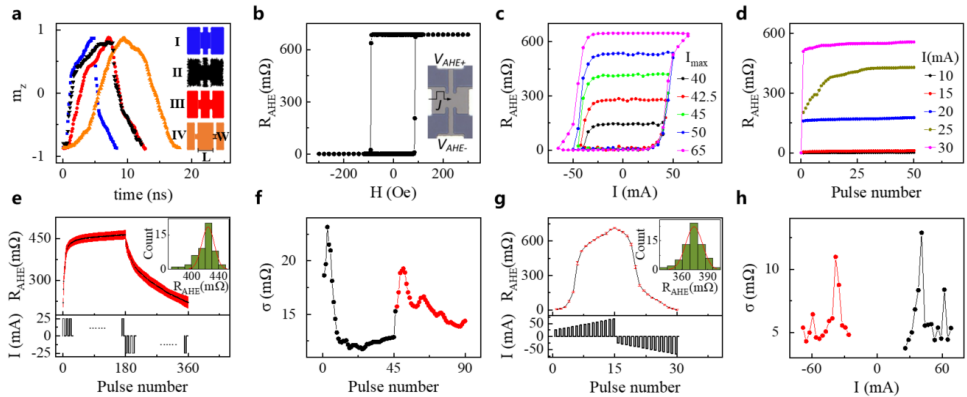

**Figure 2.** (a) Simulations of SOT-induced magnetization switching in Hall bars with varied shapes in cross regions. (b) Measurement of AHE resistance ($R_{AHE}$) of the Ta(3.0)/Pt(5.0)/Co(1.15)/SiN(7.0) multilayer. Inset:

Hall bar morphology. (c) $R_{AHE}$ Loops versus current $I$ at different maximum $I$ values. (d) Variation of $R_{AHE}$ with pulse number under various $I$ values. (e) Repeated $R_{AHE}$ measurements versus pulse number (bottom) at a fixed current amplitude ($I = 25$ mA) for 50 repetitions. The black solid curve represents the mean $R_{AHE}$, and red error bars indicate the standard deviation. Inset: representative $R_{AHE}$ distribution at the 15$^{th}$ pulse. (f) Standard deviation as a function of pulse numbers in the LTP (black) and LTD (red) procedures. (g) Repeated $R_{AHE}$ measurements versus pulse number (bottom) at varying current amplitudes for 50 repetitions. The black solid curve represents the mean $R_{AHE}$, and red error bars indicate the standard deviation. Inset: representative $R_{AHE}$ distribution at $I = -38$ mA. (h) Standard deviation as a function of current amplitude in the LTP (black) and LTD (red) procedures.

Building upon the simulation results, we fabricated an SOT Hall-bar made of the Ta(3.0)/Pt(5.0)/Co(1.15)/SiN(7.0) multilayer (The numbers in parentheses denoted layer thickness in nanometers) deposited on a Si/SiO$_2$ substrate via the magnetron-sputtering (Fig.2b) (Refer to the "Methods" section for more details). The measurement of AHE revealed that the film displays perpendicular magnetic anisotropy with a coercivity of approximately 100 Oe (Fig. 2b). We also conducted the measurement of $R_{AHE}$ as a function of current ($I$) at different maximum $I$. Here we exhibit the variation of $R_{AHE}$ ($\Delta R_{AHE}$) with respect to the initial data before applying the magnetic field or current. A series of stable $\Delta R_{AHE}$ values persists after removing the currents, forming the base for the nonvolatile multi-states of a memresistor (Fig. 2c). Additionally, we also collected the variation of $\Delta R_{AHE}$ with pulse number at different $I$ ranging from 10 to 30 mA (Fig. 2d). For $I$ below 20 mA, the variation of $\Delta R_{AHE}$ was minimal, indicating negligible weight updates when the triggering was below the threshold. However, at a current as high as 30 mA, $\Delta R_{AHE}$ rapidly approaches saturation following the initial pulse injection. A continuous nonlinear variation of $\Delta R_{AHE}$ occurred under 25 mA, rendering it suitable for nonlinear weight updates.

We conducted additional measurements on $\Delta R_{AHE}$ under a series of current pulses to characterize the stochasticity (Figs. 2e ~ 2h). In Figs. 2e and 2g, the black curves represent the mean values, while the red error bars denote the standard deviation ($\sigma$). Under a fixed current amplitude (25 mA) and pulse width (50 μs), the $\Delta R_{AHE}$ exhibits a Gaussian distribution with varying $\sigma$ between 10 and 30 mΩ (The inset of Fig. 2e illustrates a representative distribution of $\Delta R_{AHE}$ at the 15$^{th}$ pulse.). Notably, $\sigma$ significantly increases in the first few pulses for both LTP and LTD stages, stabilizing as

$\Delta R_{AHE}$ approaches saturation (Fig. 2f). A similar nonlinear variation in $\Delta R_{AHE}$ and non-monotonous changes in $\sigma$ were observed under a series of current pulse with varied amplitudes (Figs. 2g and 2h). However, the $\sigma$ in this case appears smaller than that under a fixed pulse amplitude (Figs. 2e and 2f).

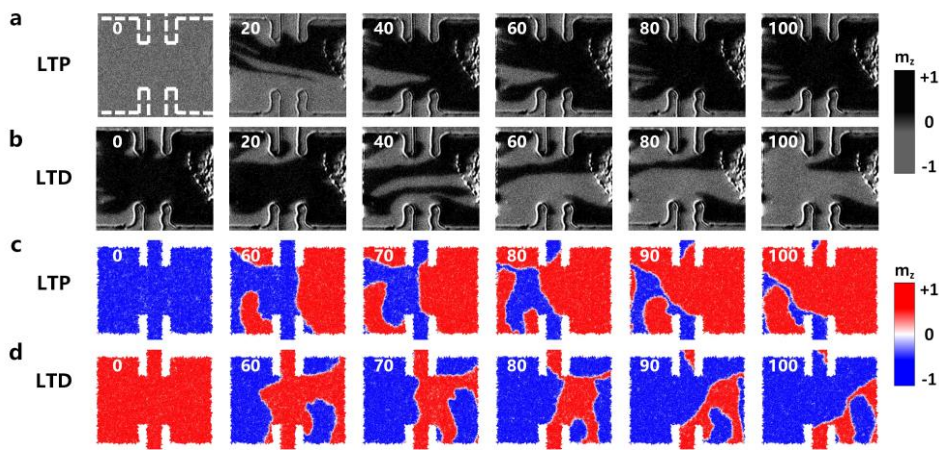

**Figure 3. (a) Visualization of DW motion under a sequence of current pulses during the LTP procedure. (b) Illustration of DW motion under a series of current pulses in the LTD process. (c) Simulation of SOT-induced DW motion during the LTP process. (d) Simulation of SOT-induced DW motion during the LTD process. The numbers labeled in the figures denote the percentage of pulses in the total number required for magnetization switching.**

To elucidate the microscopic mechanism for the nonlinear variation of $\Delta R_{AHE}$ with current pulses, we observed the SOT-induced DW motion by using the Magneto-Optical Kerr Effect (MOKE) microscope (Refer to the "Methods" section for more details). Under the initial 20 % current pulses in the LTP procedure, approximately half of the cross-region experiences rapid magnetization switching through swift DW motion immediately after depinning (Fig. 3a). Afterwards, the remaining magnetization in the cross-region was gradually switched under the subsequent 60 % pulses. In the LTD procedure, as the magnetization in the LTP stage was not fully switched, the magnetization in the cross-region easily switched during the first 40 % pulses. However, numerous additional pulses were still required to fully switch the magnetization in this region (Fig. 3b). This observed magnetization switching process aligns with the results obtained from micromagnetic simulations (Figs. 3c ~ 3d). The magnetization switching based on DW motion with varying velocity corresponds to the nonlinear variation of $\Delta R_{AHE}$ with the number of current pulses.

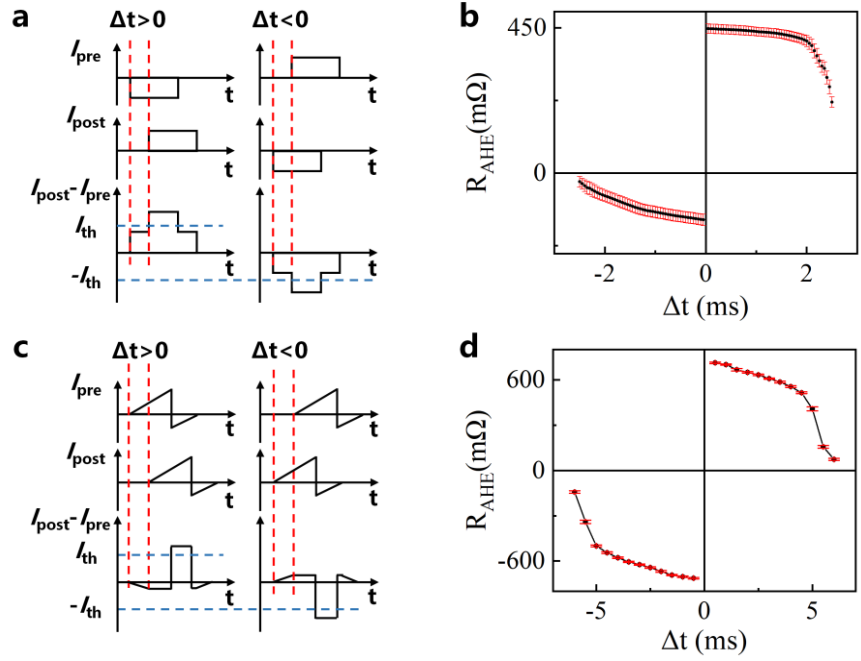

**Figure 4.** (a) Waveform of square-wave current pulses with controlled polarity designed for the pre-synaptic ($I_{pre}$) and post-synaptic ($I_{post}$) neurons, along with the total pulses ($I_{post} − I_{pre}$) influencing the synapse. $\Delta t$ is the time difference of spikes arrival between the pre-synaptic and post-synaptic neurons, and $I_{th}$ indicates the threshold current for the weight update. (b) Dependence of $R_{AHE}$ on $\Delta t$ in the scenario depicted in (a). (c) Waveform of triangle-wave current pulses applied to $I_{pre}$ and $I_{post}$, and the resulting total pulses ($I_{post} − I_{pre}$). (d) Variation of $R_{AHE}$ with respect to $\Delta t$ in the context of the triangle-wave current pulses illustrated in (c).

In an SNN, the weigh update follows the Spike-Timing-Dependent-Plasticity (STDP) rule, involving exponential weigh variation by modifying the time difference of spike arrival between the pre-synaptic and post-synaptic neurons ($\Delta t$). Typically, spiking pulses for both pre- and post-synaptic neurons can manifest as triangle waves, resulting in an effective square-wave pulse with variation in both pulse width and amplitude (Figs. 4c) [22]. Alternatively, the pulses for the pre- and post-synaptic neurons can take the form of square-wave pulses, with polarity controlled by the time sequence difference between the pre- and post-synaptic neurons (Figs. 4a) [23, 24]. By adjusting the width and amplitude of the current pulse, we verified that the variation of $\Delta R_{AHE}$ with $\Delta t$ adheres to the exponential STDP rule (Figs. 4b and 4d).

To evaluate the efficiency of UQ for the SOT synapse, we designed a stochastic SNN composed using deterministic neurons and a stochastic SOT synapse to classify the breast tumors (benign or malignant) via Python3.10 software (Fig. 5). For comparison, we also employed a deterministic SNN counterpart (with zero standard deviation in weight updates) for the same application. We mapped the $\Delta R_{AHE}$ to the weight value $\omega$ between 0 and 2 and modeled the variation of $\omega$ as a function of $\Delta t$ based on the exponential STDP rule. Meanwhile, we also fitted the standard deviation $\sigma_\omega$ of the weight update $[\sigma_\omega = (\sigma/\Delta R_{AHE})\omega]$ as a function of $\omega$.

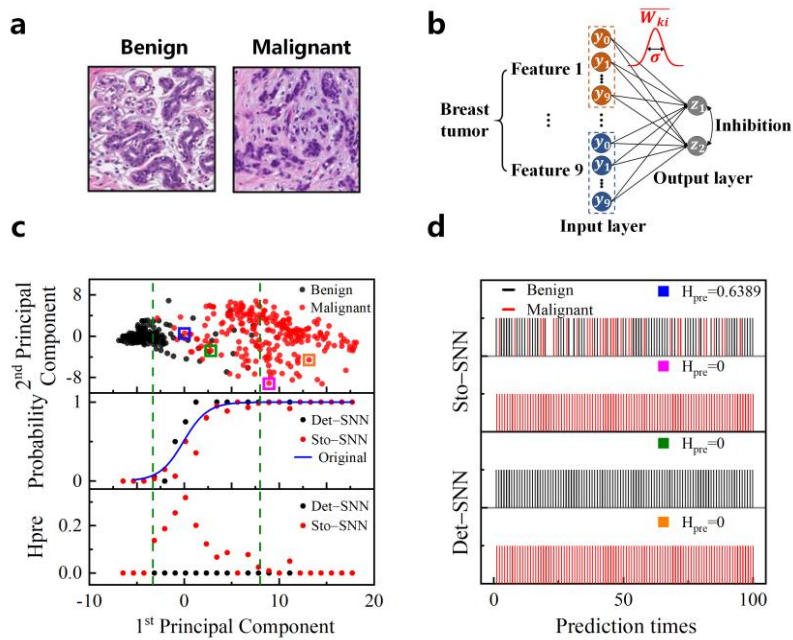

**Figure. 5 An SNN comprising stochastic DW synapses and its application in the classification of breast tumors. (a) Typical computed tomography images of benign and malignant breast tumors. (b) The architecture of the stochastic SNN designed for breast tumor classification. The SNN includes 90 input neurons representing the 9 parameterized tumor features (10 input neurons for one feature), a stochastic SOT synapse, and 2 output neurons for tumor type classification. Here we utilized the winner-take-all structure, where the firing of one output neuron inhibits the firing of the other. (c) Top: Two-dimensional distribution of breast tumor data. Middle: Average probabilities of predicting the tumor as malignant for the test data in each region. Bottom: Prediction entropy comparison between stochastic and deterministic SNNs. (d) Representative outcomes of repeated predictions for test data within specified square regions, employing both stochastic and deterministic SNNs.**

The breast tumor data, sourced from the Wisconsin Breast Cancer Data [25], comprised a dataset of 699 entries, with 399 for training and 300 for testing. Each entry detailed nine tumor features (clump thickness, uniformity of cell size, uniformity of cell shape, marginal adhesion, single epithelial cell size, bare nuclei, bland chromatin, normal nucleoli and mitoses) and 2 classes (benign or malignant). The network, featuring 90 input and 2 output neurons, encoded each tumor feature with every 10 input neurons, while each output represented a tumor type. The output neurons adhered to the leak-integer-fire principle, and the network followed the winner-take-all rule, ensuring that one neuron could not fire meanwhile the other one was active.

We initially reduced the original 9-dimensional data to 2 dimensions using the principal component analysis (PCA) [26]. Subsequently, we partitioned the 300 test data points evenly into 23 regions and visualized their distribution in a coordinate system defined by the first and second principle components (Fig. 5a). Data points located outside the region delineated by the two green dashed lines unequivocally correspond to either benign or malignant tumors. Within this region, however, the data points for the two tumor types are intermingled.

After completing the full training of the SNN, we utilized the stochastic SNN to predict the type of breast tumors for each test datum by 100 times. The prediction probabilities in each region were then averaged. The probability of classifying a tumor as benign (B) or malignant (M) can be estimated using the formula: $P(\text{B or M}) = \dfrac{N_k}{\sum_{k=1}^{2} N_k}$. Here $N_k$ represents the total spiking occurrences of the output neuron $z_k$. As illustrated in Fig. 5b, for both the deterministic and stochastic SNNs, the probabilities for classifying the tumor as malignant closely align with those calculated directly from the test data (the blue line). We further computed the overall accuracy by diving the number of correct predictions by the total 300 test data points. The stochastic SNN achieved a maximum accuracy of 95%.

Based on the results of the prediction probability, we further calculated the prediction entropy defined as $H_{pred} = -\sum_{c} P(y=c|x)\log(y=c|x)$, where $c$ indicates the type of the breast tumor, and $x$ and $y$ are the input data and the prediction result, respectively. Notably, in the intermediate region between the two dashed lines, the prediction entropy of the deterministic SNN was consistently zero. This implies that the network provides identical results across the 100 predictions, irrespective of their correctness (the bottom figure in Fig. 5d). Such uniformity could potentially lead to misdiagnose. In contrast, the stochastic SNN exhibited a non-zero prediction entropy in the middle region, peaking when the probability for classifying the tumor as malignant hovered around 50% (the upper figure in Fig. 5d). This peak signifies the highest difficulty in accurately determining the tumor type. This elevated prediction entropy serves as an indication for doctors to consider further examinations. Outside this region, however, the prediction entropy diminishes to nearly zero, indicating a high level of confidence in the predictions.

**Conclusion**

In summary, we designed and fabricated a SOT synapse, enabling nonlinear weight updates following the STDP rule through DW motion with varying velocities. The $R_{AHE}$ of the device followed a Gaussian distribution, facilitating its modeling as a stochastic synapse to introduce a noisy environment for deterministic neurons. Leveraging these stochastic synapses, we constructed a stochastic SNN and applied it to the classification of breast tumors. The network exhibited high prediction accuracy and demonstrated proficiency in quantifying the uncertainty associated with predictions.

**Acknowledgments**

The authors acknowledge financial support from the National Key Research and Development Program of China (Grant No. 2022YFE0103300), the National Natural Science Foundation of China (No. 12274119) and the Natural Science Foundation of Hubei Province (No. 2022CFA088).

**Methods**

**Device fabrication**

The Ta(3.0)/Pt(5.0)/Co(1.15)/SiN(7.0) multilayer (The numbers in parentheses represent layer thickness in nanometers) was fabricated on a Si/SiO$_2$ substrate by using an AJA-ATC-2000 magnetron sputtering system. The base pressures were better than $5\times10^{-9}$ Torr. The metallic elements Ta, Pt, and Co were deposited through the direct current sputtering at 30 W and 3-mTorr Ar gas pressure. A 7-nm thick Si$_3$N$_4$ was deposited as an isolation layer via radio frequency sputtering at 120 W and 2-mTorr Ar gas pressure. After the deposition, the multilayer was patterned into Hall bars with a 6 μm × 20 μm cross region using lithography and ion beam etching. Finally, electrodes consisting of 100 nm Al was deposited using lithography and electron beam evaporation.

**Measurement of electrical properties**

To measure current-induced magnetization switching, we assessed the AHE resistance ($R_{AHE}$) by applying pulsed current along the *x*-axis direction under a 1000-Oe external magnetic field aligned with the current direction. To characterize the stochastic behaviors of $R_{AHE}$ with current pulse, the measurements of current-induced magnetization switching were repeated over 100 times, with careful adjustments to the width and amplitude of the current pulse.

**DW motion observed by MOKE microscope**

The MOKE microscope was utilized to observe the DW motion under a train of current pulses with uniform width. Following the initialization of magnetization along the +z direction, a pulsed current with an amplitude of 90 mA and a width of 2000 μs was applied under a 100-Oe external magnetic field aligned with the current direction, and the resulting images were recorded.

**Micromagnetic Simulation**

We conducted the simulations of SOT-induced DW motion using the "Mumax3.10" software. The cross-regions of the four Hall-bars [Fig.2a] had dimensions ($L \times W$) of 60 nm × 200 nm, 60 nm × 200 nm, 110 nm × 110 nm, and 200 nm × 60 nm, respectively. The parameters employed were as follows: the cell dimension was 2 nm × 2 nm × 0.8 nm, $M_S$ (saturation magnetization) = $5.4 \times 10^5$ A·m$^{-1}$, $A$ (exchange stiffness constant) = $1 \times 10^{11}$ J·m$^{-1}$, $\alpha$ (Gilbert damping coefficient) = 0.27, $D$ (Dzyloshinskii-Moriya interaction constant) = 2.0 mJ·m$^{-2}$, $B_x$ (external magnetic field) = 0.3 T, $K_u$ (uniaxial anisotropy constant) = $8 \times 10^5$ J·m$^{-3}$, and $\theta_H$ (spin Hall angle) = 0.1. The temperature module, set to 300 K, was incorporated to introduce stochasticity. A magnetic field of 2000 Oe was applied along the *x*-axis direction, and the currents were introduced along the *x* direction, with the current densities ranging from approximately $2.5 \times 10^{11}$ to $3.5 \times 10^{12}$ A/m$^2$, adjusted based on variations in widths at different positions within the Hall bar.